# Electron Emission Energy Barriers and Stability of Sc$_2$O$_3$ with Adsorbed Ba and Ba-O

*Ryan M. Jacobs,[1] John H. Booske,*[1,2] and Dane Morgan*[1,3]*

[1]Interdisciplinary Materials Science Program, University of Wisconsin-Madison, Madison, Wisconsin 53706, USA

[2]Department of Electrical and Computer Engineering, University of Wisconsin-Madison, Madison, Wisconsin 53706, USA

[3]Department of Materials Science and Engineering, University of Wisconsin-Madison, Madison, Wisconsin 53706, USA

**ABSTRACT:** In this study we employ Density Functional Theory (DFT) methods to investigate the surface energy barrier for electron emission (surface barrier) and thermodynamic stability of Ba and Ba-O species adsorption (relative to formation of bulk BaO) under conditions of high temperature (approximately 1200 K) and low pressure (approximately 10$^{-10}$ Torr) on the low index surfaces of bixbyite Sc$_2$O$_3$. We employ both the standard Generalized Gradient Approximation (GGA) and the hybrid HSE functional to calculate accurate surface barriers from relaxed GGA structures. The role of Ba in lowering the cathode surface barrier is investigated via adsorption of atomic Ba and Ba-O dimers, where the highest simulated dimer coverage corresponds to a single monolayer film of rocksalt BaO. The change of the surface barrier of a semiconductor due to adsorption of surface species is decomposed into two parts: a surface dipole component and doping component. The dipole component is the result of charge rearrangement at the surface and is described by the electrostatic Helmholtz equation. The doping component is due to charge transfer from the surface species, which changes the Fermi level and thereby changes the surface barrier. Different initial geometries, adsorption sites, and coverages were tested for the most stable low index $Sc_2O_3$ surfaces ((011) and (111)) for both atomic Ba and Ba-O dimers. The lowest surface barrier with atomic Ba on $Sc_2O_3$ was found to be 2.12 eV and 2.04 eV for the (011) and (111) surfaces at 3 and 1 Ba atoms per surface unit cell (0.250 and 0.083 Ba per surface O), respectively. The lowest surface barrier for Ba-O on $Sc_2O_3$ was found to be 1.21 eV on (011) for a 7 Ba-O dimer-per-unit-cell coverage (0.583 dimers per surface O). Generally, we found that Ba in its atomic form on $Sc_2O_3$ surfaces is not stable relative to bulk BaO, while Ba-O dimer coverages between 3 to 7 Ba-O dimers per (011) surface unit cell (0.250 to 0.583 dimers per surface O) produce stable structures relative to bulk BaO. Ba-O dimer adsorption on $Sc_2O_3$ (111) surfaces was found to be unstable versus BaO over the full range of coverages studied. Investigation of combined n-type doping and surface dipole modification showed that their effects interact to yield a reduction less than the two contributions would yield separately.

Keywords: work function, density functional theory, oxide surface, thermionic emission, adsorption

# I. INTRODUCTION

Scandate ($Sc_2O_3$-containing) thermionic electron emission cathodes have generated substantial research interest in the past 25 years due to their experimentally observed superior properties over conventional thermionic emitters composed mainly of W and BaO. These superior properties include higher emitted current densities, lower operating temperatures, and a high resistance to chemical contamination.[1] A large supply of emitted electrons is crucial to generating the dense electron beam necessary for high power microwave (HPM) and mmw-to-THz vacuum electronic devices (VED's).[2,3] These devices are critical components in important infrastructure technologies including radar, global communications, industrial manufacturing, scientific research and military defense systems.[2]

The high magnitude of experimentally measured emitted current densities, greater than 100 A/cm$^2$, in scandate cathodes can be understood from the classic Richardson-Laue-Dushman equation[4-6] for electron emission (below the space charge limit) from a thermionic source:

$$J(T,\Phi) = AT^2 exp\left(\frac{-\Phi}{k_B T}\right), \qquad (1)$$

where $J$ is the emitted current density, $T$ is absolute temperature, $k_B$ is the Boltzmann constant and $\Phi$ is the work function. For fixed values of $A$ and $T$, the emitted current is exponentially sensitive to small changes in $\Phi$, with reductions in $\Phi$ yielding increases in $J$. For an ideal model the constant $A$ will be written as $A_{id}$ and is known as Richardson's constant, equal to 120 A/cm$^2$K$^2$. Although $A_{id}$ is comprised of only fundamental constants, the actual value of $A$ varies widely in experimental thermionic emission data due to the patch effect (where different terminating surfaces in polycrystalline samples have different work functions), or a temperature dependence of the work function due to thermal expansion of the crystal lattice.[7] Therefore, $A$ is commonly denoted the "effective" Richardson constant, as it is material-specific and sensitive to the surface conditions of the emitter.

Recent experimental results of thin film scandate cathodes with BaO present report a measured effective work function of 1.41 eV,[8,9] while some impregnated scandate cathode results yield measured effective work functions as low as 1.14 eV,[10] and a laser ablation deposited scandate top layer cathode with a measured effective work function of 1.16 eV has been made.[11] Scandate doped dispenser cathodes have been fabricated and found to have an effective work function in the range of 1.3-1.4 eV.[12] There have also been recent experimental reports of emission characteristics[13-15] and life tests[13] of nanosized scandate dispenser cathodes. The experimental work functions quoted here are deemed "effective" because they are obtained from fitting the work function indirectly. This is done either by fitting the Richardson-Laue-Dushman equation[4-6] (Eq. (1)) to data of current density versus temperature and allowing the Richardson constant and work function to be adjustable parameters (Richardson line plot), or obtaining a Miram curve and constructing a practical work function distribution (PWFD) curve by fitting the ideal form of Eq (1) to the Miram curve data.[16] In the latter case, the effective work function is the value of the PWFD at 50% normalized emission current density. In Refs. 10 and 11, which obtain exceptionally low effective work functions, the Richardson line method is used. It is commented in Ref. 11 that when the Richardson constant is readjusted to its ideal value (used as the standard to compare to metallic emitters, see Eq.(1)), the effective work function is 1.42 eV, not 1.16 eV, which is consistent with other values obtained from the Miram curve method  A summary of these experimentally reported work functions is shown below in Table 1.

**Table 1.** Summary of experimental effective work functions for various scandate cathodes.

| Reference Number | Effective work function (eV) | Fitting method |
|---|---|---|
| 9 | 1.41 | Richardson line plot |
| 10 | 1.14 | Richardson line plot |
| 11 | 1.16 (1.42) | Richardson line plot (Richardson line plot with $A = A_{id}$) |
| 12 | 1.3-1.4 | Miram curve |

The notable work function reduction of 0.5-0.7 eV in scandate cathodes over conventional thermionic emitters enables scandate cathodes to generate pulsed current densities orders of magnitude larger (100 A/cm$^2$ to 720 A/cm$^2$)[10,11] than cathodes not containing $Sc_2O_3$ at the same temperature (0.1-1 A/cm$^2$).[11] For applications which require only relatively lower current densities of order 1 A/cm$^2$, scandate cathodes still prove useful because they can provide the necessary current densities at much lower operating temperatures than non-scandate cathodes, thus lowering the input power necessary to operate the cathode. As we will focus in this study on the work function of scandate materials, here we give a short introduction to how it is defined and its behavior understood in the context of electron emitter cathodes.

The work function is defined as

$$\Phi = E_{vac} - E_{Fermi} \ , \quad (2)$$

where $E_{Fermi}$ is the Fermi energy (electron chemical potential inside the solid) and $E_{Vac}$ is the energy at the vacuum level, which is the energy at which the electron has zero kinetic energy at a semi-infinite distance away from the emitting surface and the image charge restoring force on it can be considered negligible. The designation of "semi-infinite" distance is such that the distance perpendicular to the emitting surface is large enough to produce negligible interaction compared to the work function value between the electron and the emitting surface and, in addition, simultaneously small compared to the in-plane (lateral) dimension of the emitting surface. Note that even for a metallic system the image charge interaction falls below 0.1 eV for distances of more than 15 nm, so the image charge interaction becomes negligible very quickly and this condition is easy to realize in many situations. We note that at this semi-infinite distance the electron is predominantly influenced by the surface it came from, not other nearby surfaces, which means that its vacuum level can be different for different types of surfaces, even from the same material or materials with the same Fermi level. As cathode emitters have heterogeneous surfaces made of patches of locally homogeneous regions (e.g., certain surface terminations or regions of Ba coating) we will assume that emitter cathodes have multiple patches with different work functions in the following discussions. For this assumption to be valid it must be true that the relevant work function for the cathode emitter is effectively determined at a distance that meets the semi-infinite distance criteria given above. This can be easily demonstrated by the following argument. The work function in the emitter is effectively measured at what is called the "escape distance", which is the distance from the surface at which an electron in an emitter cathode system feels a net force pulling it away from the surface. The escape distance defines when an electron has been released from the surface and will be available to provide current in the cathode, and it is therefore the work function at this distance that is relevant for cathode performance. An approximation for the escape distance can be made by considering a basic image-charge restoring force on the electron and balancing it with the field pulling the electron from the cathode to the anode.[17] The latter can be estimated by assuming a typical cathode-anode voltage difference of 1000 V and cathode-anode separation of 1 cm, which yields an electron escape distance on the order of 0.1 microns. This is an upper bound as often times cathodes are tested at potentials up to 10 kV, which would give an escape distance closer to 0.01 microns. This distance is large enough that image charge interactions are negligible (approximately 0.1 eV or lower). Furthermore, since typical cathode grain sizes are 1-5 microns (or more), distinct homogeneous surface regions (patches) are expected to have lateral dimensions on this scale. The escape distance is therefore still small on the scale of the patch lateral dimensions. Therefore, the escape distance meets the criteria for being an appropriate semi-infinite distance for determining the work function of a surface. These arguments show that we can consider the cathode to be made of patches of surface each potentially with its own work function given by the standard work function definition in Eq. (1).

Figure 1 shows a schematic band diagram of the relevant energy levels for a perfect bulk semiconductor and a semiconductor near the surface. This figure includes notation and concepts that will be used in the remainder of the work. At a temperature of absolute zero for a perfect bulk crystal, $E_{Fermi}$ is located in the middle of the bandgap. Surfaces and impurities can lead to formation of states within the bandgap, within the valence and conduction bands themselves, or produce a change in the value of the bandgap. The near-surface changes to the electronic structure may persist for numerous atomic layers until the effect is screened by the bulk material (approximately the Debye length of the material).

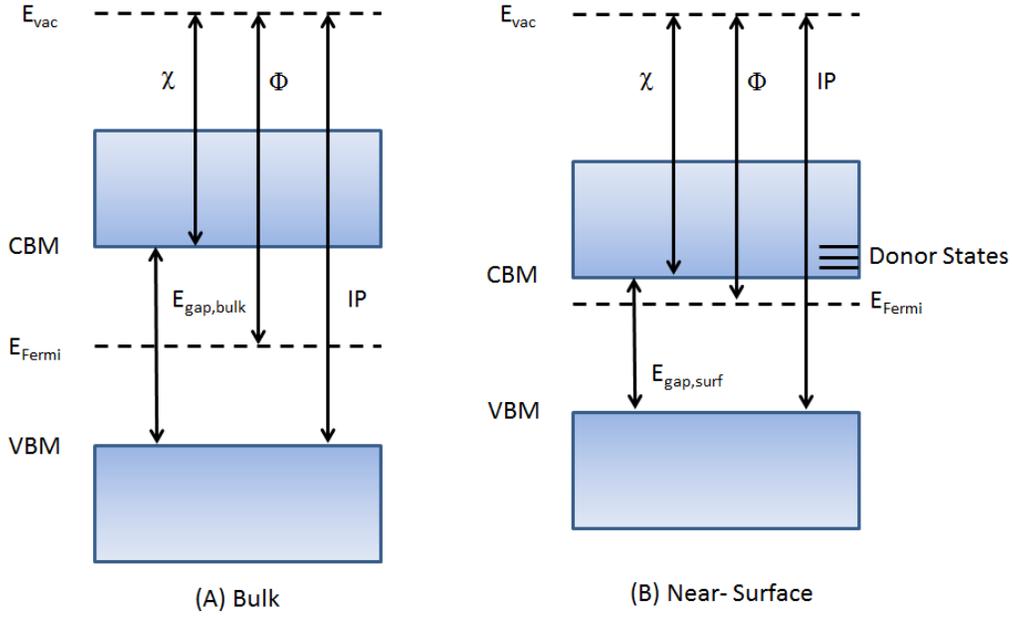

**Figure 1.** Schematic band structure of a semiconductor, (A) perfect bulk crystal at absolute zero, (B) semiconductor near the surface with donor states in the conduction band. The surface donor states ionize and the electrons relax to the conduction band minimum (CBM), raising $E_{Fermi}$ closer to the CBM. The symbols $\chi$, $\Phi$, IP and VBM stand for the electron affinity, work function, ionization potential, and valence band maximum, respectively.

Conventional thermionic emitters such as the dispenser B-type cathodes achieve a low work function by the presence of Ba-O dimers on the W terminating surface. These dimers act to form electropositive surface dipoles that lower the potential barrier for electron emission,[16,18] with work functions of about 1.9-2 eV predicted using *ab initio* methods,[19,20] in agreement with experimental values ranging from 1.9-2.1 eV.[21,22] Since the Ba-O dipole mechanism successfully explains the work function values of the W emitting surface in conventional B-type cathodes, it is plausible that the same mechanism could explain the low work functions observed in scandate cathodes. The work function change due to surface dipoles is described by the Helmholtz equation:[23]

$$\Delta\Phi_{dipole} = -\frac{e}{\varepsilon_0 A}|\vec{p}_z| \quad . \quad (3)$$

In Eq. (3), $e$ is the electronic charge, $\varepsilon_0$ the permittivity of free space, $A$ is the cross-sectional area of the emitting surface, and $|\vec{p}_z|$ is the z-component of the dipole moment between the two surfaces of the simulated surface slab (see Section IIA for details of Density Functional Theory calculations of surface slabs). In addition to creating a surface dipole it is also possible for a surface species to exchange electrons with the system and change $E_{Fermi}$. This will create another contribution to the overall change in work function, $\Delta\Phi_{Fermi}$, and we can write the total change in work function as:[24]

$$\Delta\Phi_{total} = \Delta\Phi_{dipole} + \Delta\Phi_{Fermi}, \quad (4)$$

In this equation $\Delta\Phi_{dipole}$ is given by Eq. (3) and $\Delta\Phi_{Fermi}$ is the shift of $E_{Fermi}$ between the bare surface material and the material with adsorbed species. Surface species will only significantly modify the total system $E_{Fermi}$ (through doping from surface species) when the total system density of states (DOS) near $E_{Fermi}$ is comparable to the number of electrons exchanged with the surface species and the exchange electrons can penetrate throughout the system. Such a situation is unlikely for macroscale bulk materials due to the very large number of states that may be available near $E_{Fermi}$ and electron screening mechanisms at the surface, but could occur for nanoscale thin films, and the latter are frequently used in *ab initio* studies due to limitations on system size. Furthermore, such a situation is less likely for metals, with a high DOS at $E_{Fermi}$, than for semiconductors, where the DOS at $E_{Fermi}$ can be extremely low. The consideration of both terms is therefore particularly important in very thin (nanoscale) semiconducting systems. The doping component ($\Delta\Phi_{Fermi}$) is present in nanoscale systems but is reduced to zero for macroscale materials due to screening of the limited charge available from surface species, and the entire surface barrier lowering will be due to dipole formation. Bulk dopants are well-known to alter $E_{Fermi}$, which will be captured by the $\Delta\Phi_{Fermi}$ term. In general, bulk doping can also affect the value of the surface dipole, and therefore may impact the $\Delta\Phi_{dipole}$ as well. We stress that the $\Delta\Phi_{Fermi}$ term can include contributions from both bulk doping and surface doping from adsorbed species, and that both of these doping mechanisms can occur in real cathodes. However, the surface doping contribution of $\Delta\Phi_{Fermi}$ will tend to zero in the macroscale limit, leaving only a bulk doping contribution, if bulk donors are present. We note that the computed work function modulation of metallic emitters like W in the presence of adsorbed atoms or dimers such as Ba and Ba-O are very accurately characterized by Eq. (3) without including the $\Delta\Phi_{Fermi}$ term.[1,19,20] This success shows that even for the thin slabs used in computations the surface species exist in too low quantity to alter the Fermi level in many metallic systems.

Computational studies have been previously used to explain work function modulation of metallic materials with different surface adsorbates. For example, Refs. 25-28 identify and explain the mechanisms of work function modulation of metallic substrates with adsorbed atoms or insulating ultrathin film coatings. In Ref. 25 three main contributions that affect the work function of metallic substrates with thin insulating films were identified: charge transfer, electrostatic compression, and surface relaxation. These three terms are coupled together, and constitute different components of what we refer to as the "dipole component" in this work.

It is clear from Eq. (2) that the Fermi level plays an essential role in the value of the work function. Unfortunately, the position of $E_{Fermi}$ for $Sc_2O_3$ in experimental scandate cathodes has yet to be measured. Even 99.998% pure $Sc_2O_3$ (Alfa Aesar) contains a large variety of impurity elements, which can potentially change the work function from its intrinsic value by making $Sc_2O_3$ n-type (lower work function) or p-type (higher work function). However, it is likely that the active material in scandate emitters is electron doped due to the high currents that can be achieved and the large number of electrons that are emitted.[29] A full analysis of the effects of these unknown impurities is beyond the scope of the current work. However, to capture the range of possible effects, in this study we will consider both pure and n-type doped $Sc_2O_3$.

Density Functional Theory (DFT, methods explained in the next section) calculations of metals yield accurate work functions that agree well with experiment.[20] However, by convention, in DFT calculations $E_{Fermi}$ is taken as the energy of the highest filled state, which for a perfect (no defects or impurities) semiconductor is located at the valence band maximum (VBM). Therefore, DFT calculations of energies to move electrons from a material to the vacuum level for semiconductors without defect states in the gap yield values that are actually ionization potentials (IP), not work functions. In general, the work function of a semiconductor may range between the IP (upper bound) and electron affinity ($\chi$, lower bound), where the latter is the energy to move an electron from the conduction band to the vacuum level. Given the specific physical meanings of the different types of energies associated with electron removal from a material and placement at the vacuum energy (see Fig. 1), it is useful to have a phrase that simply means the energy to move an electron from the simulated material to vacuum for a given DFT calculation. In this work we will therefore generally refer to the energy to remove an electron to the vacuum level as the "surface barrier". The surface barrier can be an ionization potential, work function, or electron affinity, depending on the situation.

There have been numerous attempts to describe the physics behind the enhanced emission of scandate cathode systems. These theories include the formation of Sc-O dipole layers on the W surface which lower the work function,[30,31] a semiconductor model in which an applied potential lowers the emission barrier near the $Sc_2O_3$ surface,[32] and formation of Ba-Sc-O surface complexes on W that reduce the work function.[20,33] There have also been claims from experiments of nanosized scandate dispenser cathodes that a thin semiconductor layer composed of Sc-Ba-O on the W surface acts to lower the work function.[15,34] Lastly, analytical models have been invoked to explain the origin of low work function via formation of a two-dimensional electron gas on the emitting surface.[35,36] In this study we explore the hypothesis that $Sc_2O_3$ with Ba-O surface complexes provide a low work function electron emitter. This hypothesis is reasonable to consider because of the general agreement that a Ba-O dipole mechanism produces low work functions in B-type cathodes,[20] experimental data on scandate cathodes which indicate the inclusion of $Sc_2O_3$ is critical for maximum work function reduction of the cathode,[9-12] and previous computational work which demonstrated that $Sc_2O_3$ itself can supply the magnitude of emitted current densities observed in experiment and thus may function as the electron emissive material in scandate cathodes.[29] We therefore consider the stability and surface barriers of different $Sc_2O_3$ surfaces, and determine the effect of adsorbed Ba and Ba-O on the $Sc_2O_3$ surface barriers to enhance the emissive capability of $Sc_2O_3$. The primary focus is on perfect $Sc_2O_3$ but some investigation of the effect of a bulk impurity (n-type) dopant on the electron emissive surface barrier of $Sc_2O_3$ is also given.

## II. COMPUTATIONAL METHODS AND THEORETICAL BACKGROUND

**A. DFT Calculation methods.** All calculations of total supercell bulk and slab energies were performed using Density Functional Theory as implemented by the Vienna *ab initio* simulation package (VASP)[37] with a plane wave basis set. The electron exchange and correlation functionals were treated with the Generalized Gradient Approximation (GGA) as parameterized by Perdew, et al., (PW91)[38,39] and projector augmented wave (PAW)-type pseudopotentials[40] for Sc, O, Ba and Li atoms. One must be cautious when calculating surface barriers of semiconducting materials using DFT and comparing with experimental work function measurements. A well-known issue with DFT is the inability to correctly place the VBM, CBM, and Fermi level with respect to the vacuum level if the LDA or GGA exchange and correlation functionals are utilized.[41] To avoid these problems, the hybrid functional of Heyd, Scuseria and Ernzerhof (HSE)[42] was used to calculate accurate bandgaps, band levels and vacuum levels for surface slab simulations. This approach was also used by Walsh and Catlow for $In_2O_3$ surface barrier calculations.[41] The bandgaps of bulk $Sc_2O_3$ from experiment, GGA and HSE calculations are 5.7-6 eV, 3.9 eV and 5.8 eV, respectively.[29,43-45] The calculated bandgap for the pristine (011) and (111) surfaces of $Sc_2O_3$ using GGA are 2.12 and 2.33 eV, respectively, and are equal to 3.47 and 3.91 eV when HSE is used. HSE was used with 25% Hartree-Fock exchange. All calculated surface barrier values reported in this work were obtained using the HSE functional. The valence electron configurations of the Sc, O, Ba and Li atoms utilized in the calculations were $3p^6 4s^2 3d^1$, $2s^2 2p^4$, $5s^2 5p^6 6s^2$ and $1s^2 2s^1$ respectively. The plane wave cutoff energy was set to 290 eV. A higher plane wave cutoff energy of 400 eV was used to check convergence of the $Sc_2O_3$ surface barriers and total energy, and it was found that they changed by no more than 0.02 eV and 1 meV/formula unit, respectively, indicating satisfactory convergence of these quantities. Reciprocal space integration in the Brillouin Zone was conducted with the Monkhorst-Pack scheme.[46] A 4x4x4 k-point mesh was used for the 1x1x1 (40 atoms) primitive cell of $Sc_2O_3$. At higher k-point densities of 5x5x5 or greater for the 40 atom cell the total energy differed by less than 1 meV/formula unit compared to the 4x4x4 k-point mesh, which is consistent with previous *ab initio* studies on bulk bixbyite $Sc_2O_3$.[29,47] For larger supercell slabs, the k-point mesh was decreased to 4x4x1, 3x2x1 and 3x3x1 for GGA calculations of the (001), (011) and (111) surface slabs, respectively. Each slab had no less than 15 Å of vacuum above the surfaces. For all calculations, care was taken to scale the k-point mesh values inversely with the supercell dimensions whenever the supercell size was varied and keep the k-point density in reciprocal space approximately constant. In order to make the computationally intensive HSE calculations more tractable they were performed with a 1x1x1 k-point mesh and without any relaxation, using the GGA coordinates. It was verified for a few cases that a static HSE calculation caused the surface barrier values to vary by no more than 0.05 eV when compared to a full relaxation using HSE. In addition, the use of the GGA k-point mesh values of 4x4x1, 3x2x1 and 3x3x1 for HSE calculations was found to produce a maximal change in the calculated surface barriers of 0.05 eV compared to using a 1x1x1 k-point mesh. All thermodynamic calculations of surface energy and Ba species adsorption energy were calculated using our GGA results. Comparing the thermodynamics of a few select HSE calculations verified that the qualitative trends in surface energy and Ba adsorption energy were the same as the GGA calculations.

When performing surface calculations, the vacuum region above the terminating surface must be thick enough to ensure vacuum level convergence for accurate surface barrier calculations. For the current study it was found that a vacuum region of 15 Å was thick enough to yield converged vacuum levels for all surface terminations. Although the thickness of our vacuum region varied on a case-by-case basis, it was verified on multiple cases that the electron vacuum energies were converged to within 0.05 eV with respect to vacuum thickness. We also explored convergence with respect to number of layers of $Sc_2O_3$ (slab thickness). We verified that the change in surface energy and surface barrier when increasing slab thickness by one layer were only 0.001 eV/Å$^2$ and 0.01 eV, respectively. Slab thicknesses were 14 (140), 11 (220), and 6 (240) ionic layers (atoms) for the (001), (011), and (111) surfaces, respectively. The cross-sectional areas for each surface unit cell differ, and are equal to $a \times a$ for the (001), $a \times \sqrt{2} a$ for (011) and $\sqrt{2} a \times \sqrt{2} a$ for the (111), where $a$ is the fully relaxed $Sc_2O_3$ lattice constant, equal to 9.871 Å.[29] For the simulations involving Ba atom adsorption, a full monolayer of Ba is defined as 6 Ba per surface unit cell for both the (011) and (111) surfaces. For Ba-O dimer adsorption, a full monolayer is defined as 8 Ba-O per surface unit cell for the (011) surface and 9 Ba-O per surface unit cell for the (111) surface. The definitions of what constitutes one monolayer resulted from initial tests of the preferred bonding arrangements of Ba and Ba-O on these surfaces, as well as determining at what coverage the Ba species fail to remain on the surface or fail to bond with the surface based on Bader charge analysis. More information on the different Ba and Ba-O coverages can be found in Sections IIIB and IIIC. In all cases, a dipole correction was implemented in VASP to ensure vacuum level convergence. This dipole is oriented perpendicular to the surface termination of interest. Because there were two surfaces present in each slab calculation, we relaxed the first three layers of each side of the surface slab for the (001) and (011) surfaces and the first two layers of each side for the (111) surface, and froze the remainder of the slab layers to be bulk $Sc_2O_3$.

The surface barrier of any surface can be determined by calculating the electron energy in the vacuum and obtaining $E_{Fermi}$ directly from the VASP calculations. The vacuum energy is determined by plotting the planar averaged electrostatic potential as a function of the coordinate normal to the terminating surface of interest, and is equal to the converged potential value normal to the surface in the middle of the vacuum region.[20] It is worth noting there are other computational methods besides DFT which have been used to calculate band energies and electron removal energies with respect to an absolute energy reference. Some of these include: a semiempirical method utilizing atomic ionization potentials and the Madelung energy of the crystal, and a combined quantum and molecular mechanical model. A summary of these methods (including DFT calculations) can be found in Ref. 48.

Every surface considered in our study was stoichiometric (no sub-oxides of $Sc_2O_3$ were studied), however very few surfaces were symmetric. For example, when calculating the surface barrier of a $Sc_2O_3$ surface with an adsorbed Ba-O dimer, only one surface termination has the Ba-O dimer present and the other surface is bare $Sc_2O_3$. The surface barriers for both the bare surface and the surface with the Ba-O dimer can be calculated and the effect of Ba-O dimer on the $Sc_2O_3$ surface barrier can be directly studied by independently calculating the terms of Eq. (4). Figure 2 provides a schematic illustration of how to calculate the $\Delta\Phi_{Fermi}$ and $\Delta\Phi_{dipole}$ terms. $\Delta\Phi_{Fermi}$ is calculated from the difference of the surface barrier of the uncoated surface for the system containing the Ba-O dimer with the reference surface barrier, given by Eq. (5). The reference surface is the equivalent bare $Sc_2O_3$ slab where no Ba-O is present on either side of the surface slab. $\Delta\Phi_{Fermi}$ is obtained with the equation

$$\Delta\Phi_{Fermi} = \Phi_1 - \Phi_{1,ref} = (E_{vac,1} - E_{vac,1,ref}) - (E_{Fermi,1} - E_{Fermi,1,ref}) \quad , \quad (5)$$

where $E_{vac,1}$ and $E_{vac,1,ref}$ are the vacuum levels of the $Sc_2O_3$ slab with Ba species and the reference slab, respectively, and analogously for the Fermi energies $E_{Fermi}$ and $E_{Fermi,ref}$. This equation aligns the vacuum levels of the bare surface slab with the surface slab containing the Ba species, and calculates the difference of the Fermi energies of the two slabs. The dipole contribution $\Delta\Phi_{dipole}$ can be directly obtained from the VASP dipole calculation and Eq. (3). Equivalently, from Fig. 2, the dipole contribution can be calculated with $\Delta\Phi_{dipole} = \Phi_2 - \Phi_1$. Note once again that the change in the dipole moment is between the two surfaces of the simulation slab, as seen in Fig. 2.

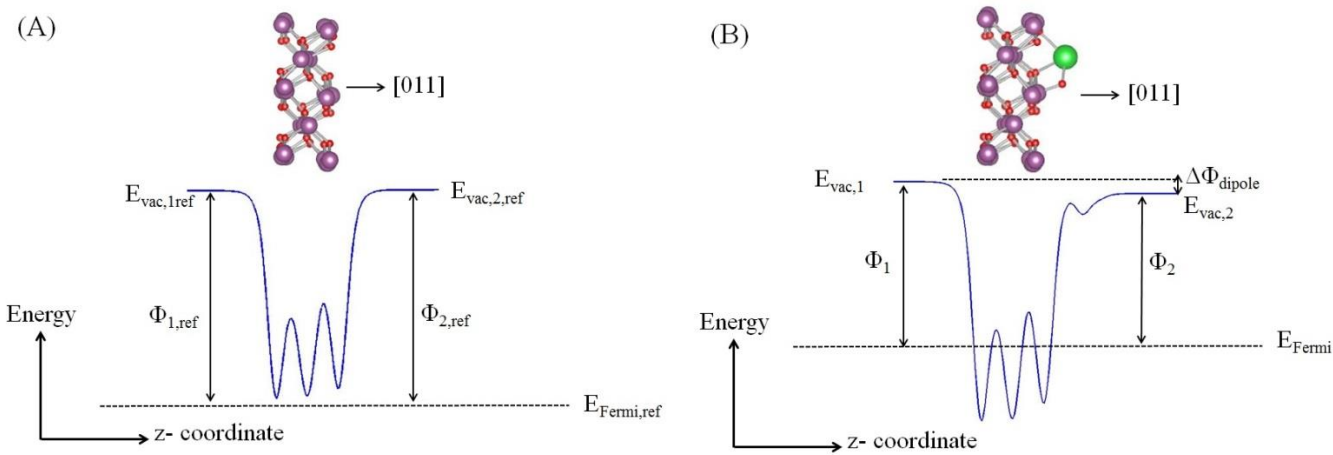

**Figure 2.** Schematic illustration showing relevant quantities to calculate the $\Delta\Phi_{dipole}$ and $\Delta\Phi_{Fermi}$ terms for a 3 layer $Sc_2O_3$ (011) surface (A) bare reference surface and (B) with a Ba-O dimer. The energy plot is of the planar (x-y) averaged electrostatic potential through the z-coordinate of the supercell. All variables are as defined in the main text. To find $\Delta\Phi_{Fermi}$ from this schematic, one must subtract $\Phi_1 - \Phi_{1,ref}$.

It is worth commenting on the expected differences in surface barrier between the calculations performed here for small nanoscale slabs with DFT and what should manifest in nature for macroscopically thick materials. When using DFT, there is an inherit computational limitation and one can only study "thin" slabs that, although they display converged surface barriers with respect to thickness, still have characteristics of a nanoscale system. This is most notable when calculating the $\Delta\Phi_{Fermi}$ term of Eq. (4) when surface dopants are present. In a DFT calculation, doping one surface of the supercell not only lowers the surface barrier of the surface containing the dopant, but also the complementary surface on the other side of the supercell that is undoped. This is because the Fermi level of the entire system has been raised by the presence of the surface donor. However, for a macroscopically thick slab, which could contain thousands of planes or more, this change in Fermi level would not occur. Instead, the donated electrons from the surface adsorbate would be screened approximately over the Debye length of the material. Therefore, for simulated systems significantly thicker than the Debye length one would expect the surface barrier of the undoped surface to approach that of the bare reference slab calculation (i.e. the charge is screened and the undoped surface does not feel the effect of the surface adsorbate). In this study, the limited size of our slabs gives a representation of surface doping effects for a nanoscale material whose thickness is such that the donated surface charge cannot be fully screened.

To illustrate this point, Figure 3 depicts schematic representations of the band diagrams near the surfaces of a semiconductor slab in the nanoscale and macroscale limits. The shaded parts of the band diagram represent filled states (valence band, "VB" in Fig. 3A) and unshaded regions are empty states (conduction band, "CB" in Fig. 3A). The black dotted lines on the band diagrams denote the position of the DFT Fermi levels for each case. Although the concepts discussed here are presented in a general way, this serves as a model to describe Ba surface doping on the $Sc_2O_3$ slabs discussed in this work. In Fig. 3, the energies given represent hypothetical surface emission barrier values. The black dotted lines that cut through the slab structures indicate the region the band diagram plot represents. Figure 3(A) is the case of no surface doping, and as one would expect the surface barrier for each surface of the slab is identical. In Figure 3(B), a surface donor is present on the top surface. The nanoscale slab exhibits a reduced surface barrier on both surfaces due to incomplete screening of the donated electrons (i.e. the nanoscale slab is thinner than its Debye length). However, for the macroscale slab, the surface barrier is only reduced locally around the surface donor, and the material recovers its undoped surface barrier value of 4.5 eV at a distance such that complete screening of the donated electrons occurs.

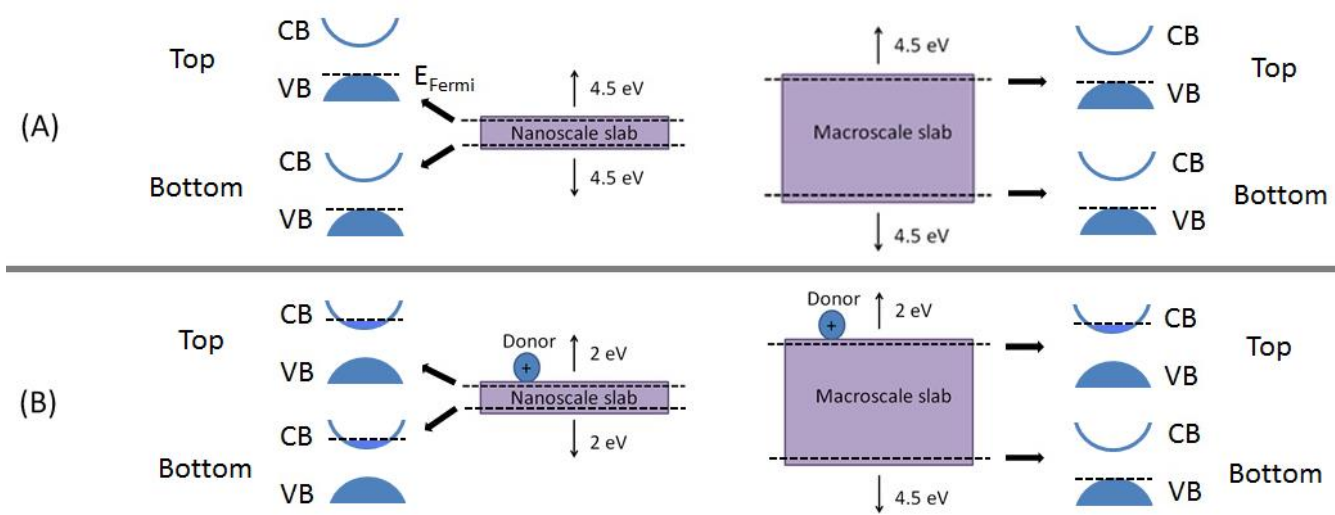

**Figure 3.** Schematic near-surface band diagram plots for nanoscale and macroscale slabs for (A) pristine surface, no surface doping and (B) with a surface electron donor present. The energy values are hypothetical surface emission barrier values. In the band diagram plots, shaded regions denote filled states and unshaded regions are empty states, and the black dashed lines represent the position of the VASP Fermi level. We use the abbreviations VB for valence band and CB for conduction band. The black dashed lines that cut through the slab drawings indicate the region of the material the band diagram plot represents, with arrows as a guide for the eye. In (B) the surface donor states are localized to the top atomic layer of both the nanoscale and macroscale slabs. The surface barrier is lowered for both surfaces in the nanoscale slab due to the material not completely screening the donated charge, however the macroscale slab recovers its undoped surface barrier value because the donated electrons are completely screened.

Our calculations on doped surfaces that are terminated with Ba or Ba-O are accurate because we have demonstrated their calculated surface barrier values converge with respect to slab thickness. However, in the limit of a macroscopic slab, the $\Delta\Phi_{Fermi}$ component of Eq. (4) will vanish due to charge screening, and the surface barrier of the undoped surface will be recovered. In this macroscopic limit, the surface barrier lowering due to adsorbates will be completely due to dipole effects. The primary reason that the surface barrier lowering is not due entirely to dipole effects in our nanoscale slabs as compared to previous studies on metals is because of the longer screening lengths in oxide materials compared to metals. As a point of comparison, a Ba or Ba-O adsorbate on a W (001) surface displays a surface barrier lowering due entirely to dipole effects even when the W layer is very thin (only a few atomic layers), due to the very short screening lengths in metals. Because of the long screening lengths for $Sc_2O_3$ the reported values of the surface barriers for the corresponding bare surfaces of our nanoscale slabs when Ba or Ba-O adsorbates are present are only correct for nanoscale thin films and may have a strong thickness dependence. Similarly, the component of the surface barriers of the surfaces containing the Ba or Ba-O due to $\Delta\Phi_{Fermi}$ is only correct for nanoscale thin films and may also have a strong thickness dependence, going to zero for thick slabs. However, the total surface barriers of the surfaces containing the Ba or Ba-O adsorbates are representative of *both* nanoscale and macroscale materials. Figure 4 shows a schematic of the slab thickness dependence of the $\Delta\Phi_{Fermi}$ term and Fermi level as a result of surface electron donors. The graphic in the upper part of Fig. 4 depicts a surface slab with two surfaces: $\Phi_1$ is the surface barrier for the corresponding bare surface and $\Phi_2$ is the surface barrier of the surface containing the electron donor. For small slab thickness (nanoscale regime), the material cannot fully screen the donated charge and thus there is a sizeable $\Delta\Phi_{Fermi}$ contribution. As the material becomes thick enough such that the Debye length is reached, the donated charge becomes fully screened and the bare surface becomes unaffected by the now distant surface donors. In this way, the $\Delta\Phi_{Fermi}$ term approaches zero, $\Phi_1$ approaches its reference value and the surface barrier lowering is completely due to dipole formation.

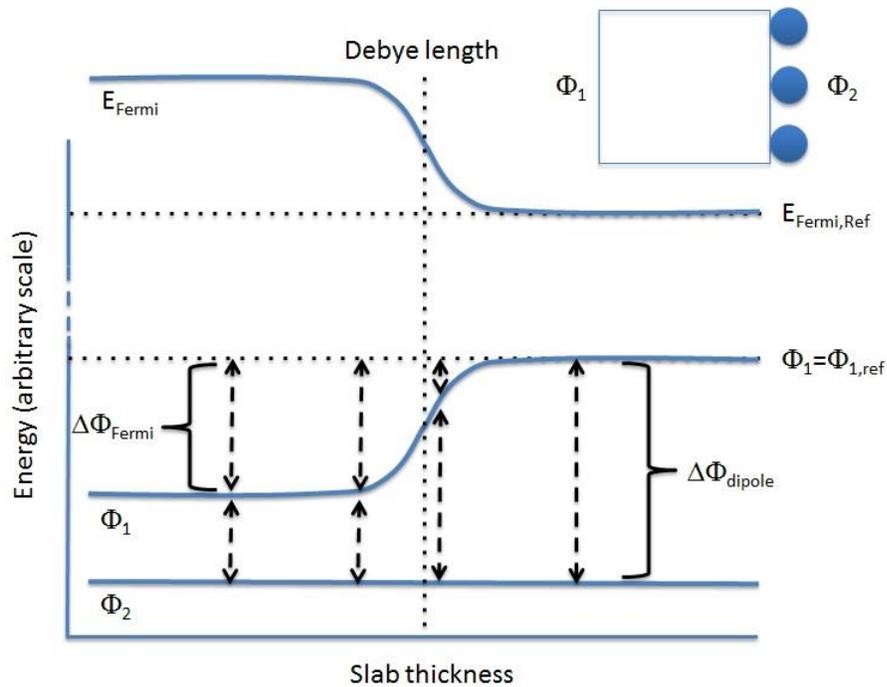

**Figure 4.** Schematic representation of the thickness dependence of $E_{Fermi}$ and $\Delta\Phi_{Fermi}$ when surface electron donors are present. The graphic depicts a system with two surfaces: a bare surface with surface barrier $\Phi_1$ and a surface containing electron donors with surface barrier $\Phi_2$. $\Delta\Phi_{Fermi}$ and $\Delta\Phi_{dipole}$ are defined in the text and values labeled "ref" refer to the the macroscale values of the system. For nanoscale slabs, the lowering of the surface barrier has sizeable contributions from both dipoles and electron doping. After the Debye length is surpassed, the donated charge is completely screened and the surface barrier lowering is completely dominated by surface dipoles.

**B. Surface stability and Ba species adsorption energy.** The surface slabs used in this study all had either a (001), (011), or (111) termination, as these low-index terminations tend to be the most thermodynamically stable by virtue of fewer broken bonds per unit surface area. The stability of a surface (including any species adsorbed on it) is characterized by the value of its surface energy, which can be directly calculated from DFT total energies and the chemical potentials of any adsorbed species. For a slab of a general binary oxide $M_mO_n$, the average surface energy $\gamma$ is:[49]

$$\gamma = \frac{1}{2A_s}\left(E_{slab}^{M_mO_n} - \frac{N_M}{m}E_{form.unit}^{M_mO_n} - \sum_{i=1}^{n} x_i \mu_i\right), \quad (6)$$

where $A_s$ is the surface area of one terminating surface, $E_{slab}^{M_mO_n}$ is the total energy of the surface slab, $N_M$ is the number of $M$ atoms present in the surface slab, $m$ is the stoichiometry of the metal species, $E_{form.unit}^{M_mO_n}$ is the energy for one formula unit of the $M_mO_n$ oxide in its bulk form, $\mu_i$ is the chemical potential of the $i^{th}$ adsorbed species (in eV/atom) and $x_i$ is the number of atoms of the $i^{th}$ adsorbed species. The factor of 2 is present because there are two surfaces with the same area $A_s$ in every supercell calculation. The first two terms in the parentheses represent the energy difference between the surface slab and the equivalent amount of material in its bulk form. This surface energy is the average of the surfaces exposed, which may not be the same type of surface.

The thermodynamics of Ba and O adsorption on $Sc_2O_3$ surfaces is governed by the chemical potential of Ba and O in the cathode operating environment (T approximately 1200 K, P approximately $10^{-10}$ Torr). Ba metal is not stable at high temperature and any significant oxygen partial pressure, therefore Ba atoms are assumed to originate from bulk BaO (rocksalt structure), while O atoms originate from a reservoir of $O_2$ gas. The chemical potential $\mu_{Ba}^{BaO}$ of Ba atoms in this environment is a function of the cohesive energy per formula unit $E_{BaO}^{form.\ unit}$ of BaO and the chemical potential of oxygen $\mu_O^0$ as:

$$\mu_{Ba}^{BaO} = E_{BaO}^{form.\ unit} - \mu_O^0 . \qquad (7)$$

The O chemical potential can be calculated by using a combination of DFT total energies and experimental thermodynamic data for $O_2$ gas at the relevant reference state.[50] It takes the form:[29,51]

$$\mu_O^0 = \frac{1}{2}\left[E_{O_2}^{DFT} + \Delta h_{O_2}^0 + H(T,P^0) - H(T^0,P^0) - TS(T,P^0) + kT\ln\left(\frac{P}{P^0}\right) - \left(G_{O_2}^{s,vib}(T) - H_{O_2}^{s,vib}(T^0)\right)\right] \qquad (8)$$

where $E_{O_2}^{DFT}$ is the DFT calculated energy of an isolated, spin polarized $O_2$ gas molecule, $\Delta h_{O_2}^0$ is a numerical correction term of the energy of oxygen in $O_2$ molecules versus a solid (this includes a correction for the enthalpy of the solid at T=0 relative to oxygen in the gas phase at T=298 K, thermodynamic contributions to the enthalpy at T=298 K from the solid phase oxygen and a correction of $O_2$ overbinding from DFT calculations), $H(T^0,P^0)$ and $H(T,P^0)$ are the gas enthalpy values at standard and general temperatures $T^0$ and $T$, respectively, $S(T,P^0)$ is the gas entropy, and the logarithmic term is the adjustment of the chemical potential for arbitrary pressure. The final terms in Eq. (8), $G_{O_2}^{s,vib}(T)$ and $H_{O_2}^{s,vib}(T^0)$, shift the value of $\mu_O^0$ to account for solid phase vibrations, which are approximated with an Einstein model with an Einstein temperature of 500 K following Refs. 20,51.

Since BaO is more stable than Ba metal in the cathode environment, we expect our modeled system to satisfy the condition that BaO is stable against the loss of O, i.e.,

$$\mu_{Ba}^{BaO} \leq \mu_{Ba}^0 , \qquad (9)$$

where $\mu_{Ba}^0$ is the chemical potential of Ba in Ba metal. To confirm that this condition is satisfied for our modeled system, we consider the Ba chemical potential value appropriate for cathode operating conditions of T

= 1200 K and P = $10^{-10}$ Torr to evaluate the stability of $Sc_2O_3$ surfaces containing adsorbed Ba species, which is given by Eq. (7) where $\mu_O^0$ is evaluated at T = 1200 K and P = $10^{-10}$ Torr. Performing this calculation yields $\mu_{Ba}^{BaO}$ = -6.064 eV/Ba, $\mu_O^0$ = -5.886 eV/O, and $\mu_{Ba}^0$ = -1.923 eV/Ba, and Eq. (9) is indeed satisfied.

The adsorption energy $E_{ads}$ of a Ba-containing species to a surface is given by:

$$E_{ads} = \frac{1}{n}\left(E_{slab} - E_{slab}^{bare} - \sum_i x_i \mu_i \right), \quad (10)$$

where $n$ is the number of adsorbed species, $E_{slab}$ is the total energy of the surface slab containing the adsorbed species, $E_{slab}^{bare}$ is the total energy of the bare surface slab, $x_i$ is the number of atoms of type $i$, and $\mu_i$ is the chemical potential of atom $i$. The chemical potentials are taken from Eq (7). As a function of surface energies prior to ($\gamma_{bare}$) and after ($\gamma_{ads}$) adsorption, the adsorption energy is expressed as:

$$E_{ads} = \frac{2A_s}{n}(\gamma_{ads} - \gamma_{bare}). \quad (11)$$

Within this formalism, one can see that for $\gamma_{ads} < \gamma_{bare}$, $E_{ads} < 0$ and the adsorbed species on the surface is more stable (in a relative sense) on the surface compared to its bulk reference state. Experimentally, Ba will evaporate from a scandate cathode over time and is thus thermodynamically unstable. This evaporation occurs for a W-BaO cathode as well, and is the primary reason why Ba must be replenished on the surface to maintain plentiful electron emission. While the long-term instability of Ba on the surface is clear, it is the transient forms Ba takes while on the surface that are critical for understanding the work function. Therefore, for this study, we calculate adsorption energies with Eq. (10) (equivalently with Eq. (11)) versus BaO as a proxy for which Ba surface structures are stable relative to others. More stable Ba arrangements will tend to reside on the surface for a longer period of time prior to evaporating and will tend to be more common, making them the most likely candidates to impact the work function.

## III. RESULTS AND DISCUSSION

**A. Bare Bixbyite Sc$_2$O$_3$ surfaces.** The structures of the fully relaxed low index Sc$_2$O$_3$ surfaces are shown in Fig. 5. We first consider the surfaces within Tasker's[52] framework for analyzing polar surfaces. The (001) surface consists of alternating layers of Sc and O atoms in the [001] direction, therefore terminations of this type produce polar surfaces, with one unique O termination and two unique Sc terminations. The layer arrangement of alternating positive and negative ions indicative of the (001) surface obeys Tasker's definition of a Type 3 polar surface. The stacking of the (011) surface contains stoichiometric amounts of both Sc and O. Therefore terminations of this type are nonpolar. The (011) surface is a Type 1 surface with the Sc and O ions alternating in the plane of the terminating surface, producing an overall nonpolar surface. The (111) can be regarded as a Type 2 surface, featuring a total dipolar repeat unit of Sc and O that makes the surface nonpolar.[52,53]

Due to the polarity of the (001) surfaces, the mean electric field in the material is nonzero, leading to a divergence of the surface energy with respect to slab thickness if the surface charge is uncompensated.[52] These polar terminations are typically very unstable, and to become more stable they require a compensation of the polarity by a process that in general may cost a lot of energy. Some compensation mechanisms include the formation of regions that are off-stoichiometric, a physical surface reconstruction, or movement of charge. Insulating oxides with polar surfaces typically exhibit poor convergence of surface barrier and surface energy values unless a large number of slab layers are used. Thick slabs are needed to compensate the polarity because the system can only move charge to compensate the surface polarity and eliminate the internal electric field once a certain slab thickness is attained, and it is not possible for this compensation to occur if the slab is too thin.[54] Following the procedure in Ref. 41 for In$_2$O$_3$ (001) surfaces (In$_2$O$_3$ also has the bixbyite crystal structure), we perform a basic (001) reconstruction by relocating one half of the O atoms on the O-terminated (001) face to the neighboring Sc-terminated (001) face, thus compensating the surface polarity. The idea for this procedure first stemmed from simulations on crystal morphology of NiO (rocksalt structure),[55] which possesses Type 3 polar surfaces of the {111} termination family, and also from STM studies of UO$_2$ (fluorite structure), which contain Type 3 polar surfaces of the {001}[27,28] termination family and were hypothesized to undergo the same type of surface reconstruction performed here.[56]

Table 2 contains the surface energy and surface emission barrier for each possible termination of the (001), (011) and (111) $Sc_2O_3$ surfaces. Based on our calculations, the order of stability (from most to least stable) is: $\gamma$ (111) < $\gamma$ (011) < $\gamma$ (001). This order of stability is consistent with a first-principles surfaces study of $In_2O_3$,[41] and also agrees with experimental single crystal growth of $Sc_2O_3$, which shows that $Sc_2O_3$ crystals grow preferentially with (111) oriented facets.[41,57,58] The (001) reconstruction stabilized the original polar (001) termination by between 0.15 to 0.16 eV/Å$^2$, however this reconstructed surface is still significantly less stable than the (011) and (111) surfaces.

**Table 2.** Calculated surface energies and surface barriers for the fully relaxed bare low index $Sc_2O_3$ surfaces. Surface energies for the (001) Sc 1 and (001) Sc 2 rows differ slightly due to the different structures of the Sc terminations present. For the (001) Sc 1 and Sc 2 cases, $\Phi_1$ and $\Phi_2$ are the surface barriers of the Sc and O terminated face, respectively. These two columns are labeled with (Sc) and (O) for clarity on which surface is being considered. All surface barriers are given under p-type conditions ($E_{Fermi}$ located at the valence band maximum).

| Surface Termination | $\gamma$ (eV/Å$^2$) | $\Phi_1$ (eV) | $\Phi_2$ (eV) |
|---|---|---|---|
| (001) Sc 1 | 0.289 | 3.90 (Sc) | 4.33 (O) |
| (001) Sc 2 | 0.291 | 3.87 (Sc) | 4.35 (O) |
| (001) Sc 2 reconstructed | 0.135 | 6.49 | 6.03 |
| (011) | 0.077 | 5.16 | 5.18 |
| (111) | 0.051 | 5.92 | 5.92 |

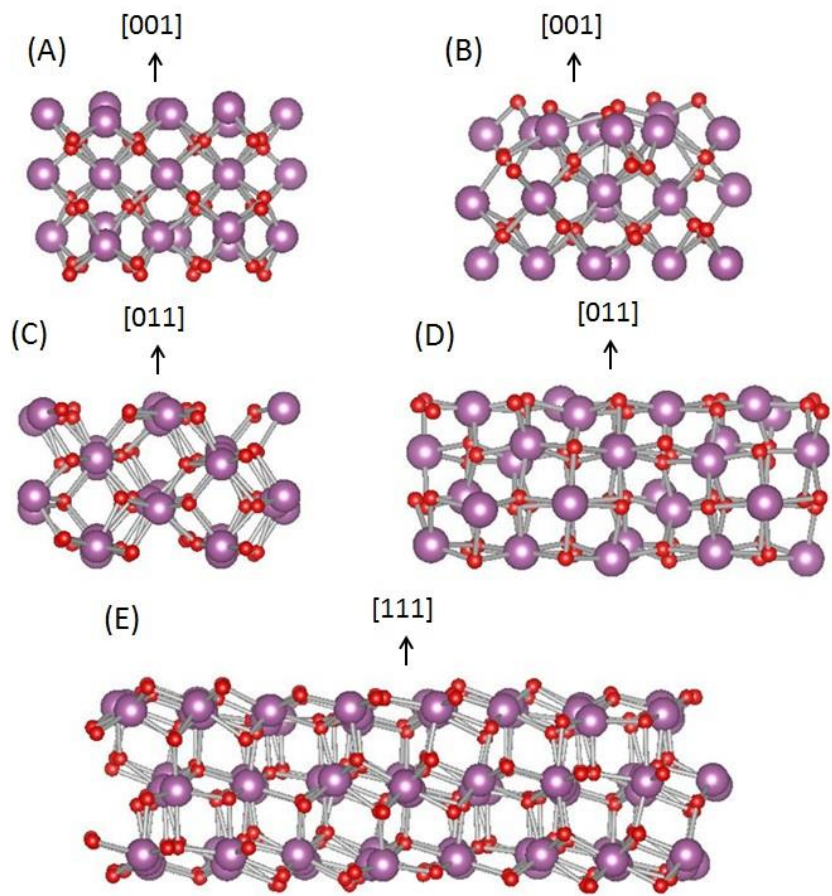

**Figure 5.** Structures of fully relaxed low index bare $Sc_2O_3$ surfaces. The Sc atoms are the large (purple) spheres while the O atoms are the smaller (red) spheres. A: The (001) surface showing one of the two possible Sc terminations (the other Sc termination is the $2^{nd}$ ionic row of Sc atoms beneath the surface), B: The reconstructed (001) surface, C: The nonpolar (011) surface, showing surface morphology. D: Same as part C, but rotated about the [011] axis by 90 degrees, E: The nonpolar (111) surface.

The (001) surfaces with either Sc termination have the lowest calculated surface barriers, ranging from 3.87-3.90 eV, while the O terminated faces have surface barriers between 4.33-4.35 eV. The lower surface barrier of the Sc termination can be attributed to the large polarity of the surface: the full layer of electropositive Sc ions forms a positively oriented dipole moment on the surface, which produces a lower overall surface barrier. A complementary argument can be made for the O terminated (001) surfaces: these have higher surface barriers, this time due to the electronegative O layer suppressing electron emission by producing an overall dipole moment which points into the material. The (001) terminations have large surface energies, are expected to be rare, and are therefore unlikely to be the dominant emitting surfaces in scandate cathodes. Consequently, further investigation of Ba species adsorption on these surfaces is not considered here. The (011) and (111) surfaces have surface barriers collectively ranging from 5.16 to 5.92 eV and due to their stability it is expected that the (011) and (111) surfaces are the most prevalent and dominate emission in real scandate cathodes.

No surface reconstructions for the (011) and (111) surfaces were considered in this study. Experimentally, there is no evidence that the $Sc_2O_3$ emitting surface should undergo any significant reconstruction, as inferred from studies of $Sc_2O_3$ films deposited on an $Al_2O_3$ (0001) substrate (2.2% lattice mismatch) with e-beam evaporation,[57] a GaN (0001) substrate (9.2% lattice mismatch) with MBE techniques, and a Si (111) substrate (9.2% lattice mismatch) also with e-beam evaporation.[59] These $Sc_2O_3$ films resulted in epitaxial, atomically flat, and uniformly thick (111) surfaces despite the large range of lattice mismatch of $Sc_2O_3$ with the underlying substrates. $Sc_2O_3$ has also been deposited on GaN (0001) with atomic layer deposition, resulting in a polycrystalline, predominantly (111) terminated $Sc_2O_3$ film also free of reconstructed surfaces.[60] There has been no consensus regarding what terminations of $Sc_2O_3$ are actually present during cathode operation or what the surface structure of $Sc_2O_3$ with adsorbed Ba looks like. Due to the lack of evidence for $Sc_2O_3$ undergoing surface reconstructions, we consider Ba adsorption mechanisms only on the non-reconstructed (011) and (111) surfaces depicted in Fig. 5.

**B. Atomic Ba adsorption on (011) and (111) $Sc_2O_3$ surfaces.** Barium is present in all types of conventional and scandate thermionic cathodes, and is the key component in work function reduction of metallic B-type cathodes composed mainly of W and BaO. During operation of impregnated cathodes, the BaO impregnant dissociates into $O_2$ gas and free Ba atoms.[61] These Ba atoms diffuse to the emitting surface, where it has been shown they form[20,22,62] electropositive dipole layers which lower the work function, thus producing higher emission current densities. More generally, Ba may exist in its atomic form, in the molecular form of Ba-O dimers, or crystalline solid form of BaO. In this section, we consider atomic Ba adsorption on different possible surface sites for the stable (011) and (111) surface terminations of $Sc_2O_3$, while the effect of Ba-O dimers and a monolayer BaO film are addressed in Section IIIC. Due to the complex arrangement of surface atoms in the bixbyite structure, different Ba configurations were used to determine which adsorption sites were the most stable for each surface termination. Once these most stable sites were found, the Ba coverage density was varied to study how the stability and surface barrier varied with increasing Ba content.

Figure 6 shows the most stable adsorption site for Ba on both the (011) and (111) surfaces. As shown, these arrangements correspond to coverage densities of $7.26 \times 10^{13}$ and $5.13 \times 10^{13}$ Ba/cm$^2$ for the (011) and (111) surfaces, respectively, which is equivalent in both cases to 1 Ba per surface unit cell. In the remainder of this work and for the sake of brevity, coverages will be given as the number of adsorbed species per surface unit cell. Dividing by the surface slab dimensions given in Section IIIA converts this specification into the conventional units of Ba/cm$^2$. For both the (011) and (111) surfaces, full Ba coverage corresponds to 6 Ba atoms per surface unit cell. For the (011) surface, Ba prefers to adopt a bridge structure with one O atom on either side of the trench-like morphology indicative of this surface. The (111) surface is composed of an array of O atoms arranged in close pairs, all of which have one unsatisfied bond. A full surface coverage of 6 Ba per unit cell for the (011) surface assumes that Ba will preferentially make more than one bond with the $Sc_2O_3$ surface O atoms (analogously with 6 Ba per unit cell for the (111) surface). A very high coverage was simulated for the (011) surface with one adsorbed Ba for each surface O atom, i.e. 12 Ba per unit cell. The resulting fully relaxed structure left only half the Ba atoms on the surface, each possessing the bridge structure shown in Fig. 5. The remaining Ba atoms desorbed from the surface and proceeded into the vacuum region away from the surface. Bader charge analysis confirmed these desorbed Ba atoms were not chemically bound to the $Sc_2O_3$ surface, as no charge transfer took place. This result demonstrates that Ba will likely bond to more than one $Sc_2O_3$ surface O atom.

The calculated values for the surface barrier, surface energy, and Ba adsorption energy for different Ba coverage densities in the stable formations depicted in Fig. 6 are summarized in Table 3. The dependence of surface barrier and surface energy on Ba coverage are plotted in Fig. 7. Ba coverages of just 1 atom per unit cell lower the surface barrier of the bare $Sc_2O_3$ surfaces from 5.16 to 2.31 eV and 5.92 to 2.04 eV for the (011) and (111), respectively. Ba coverages higher than 1 atom per unit cell further decrease the surface barrier by only 0.16-0.19 eV for the (011) surface and do not appreciably change the surface barrier of the (111) termination.

**Table 3.** Calculated surface barriers, surface energies, and Ba adsorption energies on the (011) and (111) surfaces. For the surface energies and adsorption energies the values are given using $\mu_{Ba}^{BaO}$ under cathode operating conditions. For the surfaces with nonzero Ba adsorption, $\Phi_1$ is the surface emission barrier of the Ba-containing side of the surface slab and $\Phi_2$ is the surface emission barrier of the bare side of the surface slab. The slabs, which have zero Ba coverage, serve as the reference slabs for calculating the $\Delta\Phi_{Fermi}$ term. By virtue of Ba acting as an electron donor, all surface barriers (with the exception of the reference zero coverage surfaces) are given under n-type conditions.

| Termination | Ba coverage (atoms/unit cell) | $\Phi_1$ (eV) | $\Phi_2$ (eV) | $\Delta\Phi_{dipole}$ (eV) | $\Delta\Phi_{Fermi}$ (eV) | $\gamma$ (eV/Å$^2$) | $E_{ads}^{Ba}$ (eV/Ba) |
|---|---|---|---|---|---|---|---|
| (011) | 0 (ref) | 5.16 | 5.18 | -0.02 | n/a | 0.077 | n/a |
|  | 1 | 2.31 | 1.89 | 0.45 | -3.29 | 0.092 | 3.93 |
|  | 3 | 2.12 | 1.86 | 0.26 | -3.32 | 0.124 | 4.29 |
|  | 6 | 2.14 | 1.72 | 0.42 | -3.46 | 0.162 | 3.89 |
| (111) | 0 (ref) | 5.92 | 5.92 | 0.00 | n/a | 0.051 | n/a |
|  | 1 | 2.04 | 2.31 | -0.27 | -3.61 | 0.062 | 4.17 |
|  | 2 | 2.06 | 2.22 | -0.16 | -3.70 | 0.073 | 4.21 |
|  | 4 | 2.06 | 2.21 | -0.15 | -3.71 | 0.094 | 4.22 |

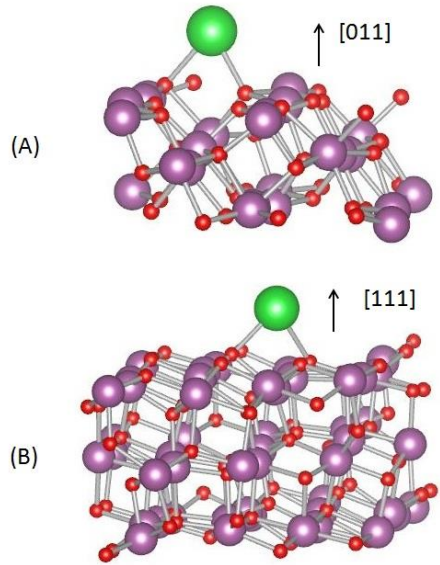

**Figure 6.** Most stable adsorption sites for Ba (largest green spheres) on the (011) and (111) $Sc_2O_3$ surfaces. A: Ba adopting a bridge configuration, bonding to two O atoms on either side of the (011) trench morphology. B: Ba bonding to two closely paired surface O atoms on the (111) surface. For visual clarity, only the top 3 ionic layers of a single surface unit cell are shown.

The shape of surface barrier vs. Ba coverage curves shown in Fig. 7 differ from what is observed for Ba on W. With Ba on W, a much deeper potential well minimum is present, and is located between 0.25 to 0.50 Ba per W(001) surface unit cell[20] (see Fig. 3 of Ref. 20). This behavior for Ba on W is due to the large dipole effect lowering the work function at low Ba coverages, with dipole depolarization occurring at higher coverages resulting in an increase in the W work function. However, atomic Ba on $Sc_2O_3$ mainly acts as an electron donor rather than a surface dipole former, as can be seen by applying the decomposition given in Eq. (4), shown in Fig. 7. For Ba on $Sc_2O_3$ the dipole effect is small and therefore very little depolarization is observed. In the case of W with Ba, the Ba donates electrons to W but the surface barrier reduction is due purely to a dipole effect because of the large density of states around $E_{Fermi}$ in W. In the case of $Sc_2O_3$, the valence band is filled, and extra states from the Ba adsorption hybridize with O states in the conduction band, so the donated electrons dope the system and raise $E_{Fermi}$ into the bottom of the conduction band. Therefore, surfaces with adsorbed Ba provide an approximate measure for the electron affinity (see Fig. 1 and surrounding discussion for definition) of the (011) and (111) $Sc_2O_3$ surfaces (plus the minor effects of surface adsorbed Ba dipoles). The reduction in surface barrier is very small beyond 1 Ba per surface unit cell because although more electrons are being doped into the $Sc_2O_3$ by the additional Ba, the density of states in the conduction band increases very rapidly. Hence, $E_{Fermi}$ remains pinned to the bottom of the conduction band and $E_{Fermi}$ cannot be significantly raised further, which implies that the emission barrier remains relatively constant. Bader charge analysis confirms that Ba donates electrons to both Sc and O states of the $Sc_2O_3$ for both the (011) and (111), however since the calculated surface emission barrier of the Ba-containing surface is higher than the corresponding bare surface, the dipole component by definition has a negative value for this case.

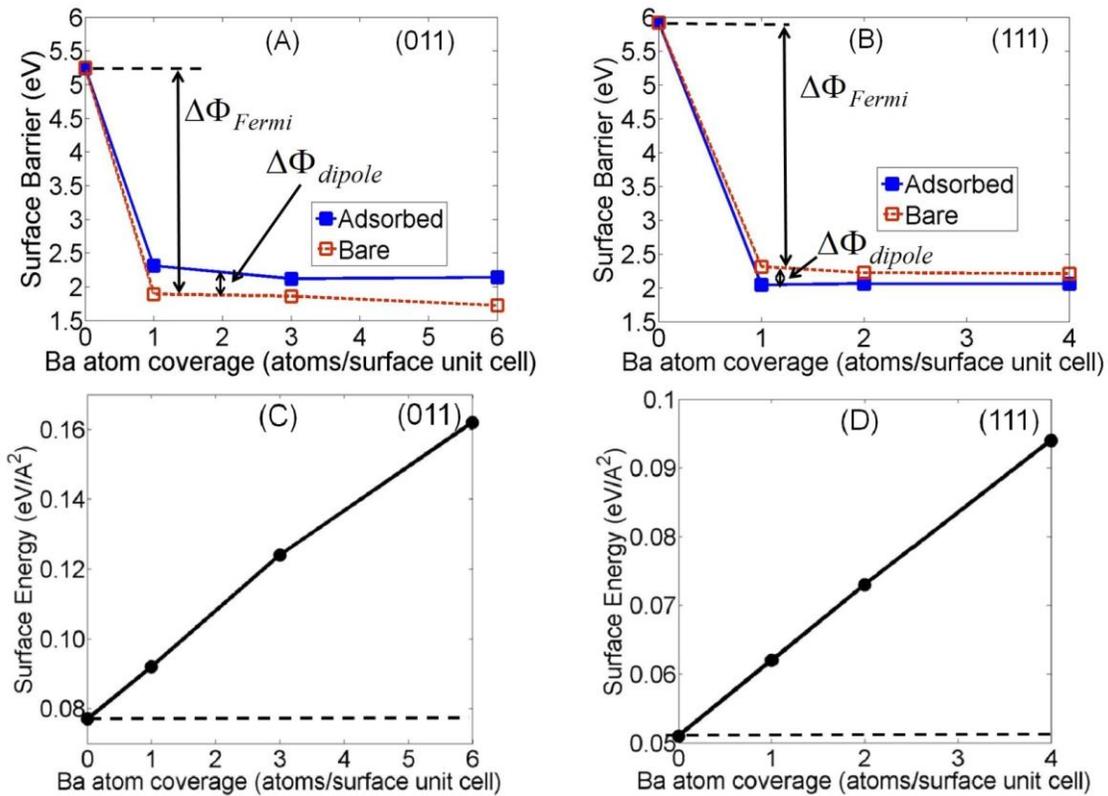

**Figure 7.** Surface barrier as a function of Ba coverage for the (011) termination (A) and (111) termination (B). Solid blue lines are surface barriers of the Ba-terminated surface while dashed red lines are the surface barriers of the opposing bare surfaces. Surface energy as a function of Ba coverage for the (011) termination (C) and

(111) termination (D) are plotted. The dashed lines in (C) and (D) mark the surface energy of the bare reference slabs for comparison.

The surface energies of all surface slabs containing adsorbed Ba are higher than their respective bare $Sc_2O_3$ slabs, resulting in positive values for the Ba adsorption energy in all cases. From Eq. (11), this implies that Ba is more stable in the BaO rocksalt structure from which it originated than on the $Sc_2O_3$ surface. Thus, Ba on $Sc_2O_3$ surfaces is not a thermodynamically stable state relative to bulk BaO. In general, the most stable Ba adsorption mechanism will be characterized by the lowest adsorption energy relative to other adsorption mechanisms, and the most stable Ba states will reside on the emitting surface for a longer period of time prior to evaporation. Since the mechanism of atomic Ba adsorption only lowers the $Sc_2O_3$ surface barrier to 2.04 eV in the best case and no atomic Ba adsorption coverages are stable versus BaO, it is not likely that atomic Ba adsorption explains the enhanced emission exhibited by scandate cathodes.

**C. Ba-O dimer adsorption on the (011) and (111) $Sc_2O_3$ surfaces**. We now further explore the effect of Ba on the properties of the (011) and (111) $Sc_2O_3$ emitting surfaces where the surface Ba is now present in the form of Ba-O dimers. The logic for pursuing Ba-O dimers is twofold: first, Ba-O dimer structures were sufficient to explain the work function lowering in conventional dispenser cathodes composed mainly of W and BaO, and second, Ba-O could be more stable than atomic Ba due to the presence of an additional bonding O, as the dimer can be thought of as a single BaO formula unit.

Ten starting arrangements of a 1 Ba-O per surface unit cell coverage on the (011) surface were simulated in order to find the most stable Ba-O adsorption geometry. This was done by locating surface O atoms for the Ba to bond to, and then situating the Ba in the unrelaxed structure such that it was closest to bonding to either a single surface O, two surface O atoms that line the trench morphology of the (011) surface, or within the trench itself near three surface O atoms. Then, the O associated with the Ba-O dimer was added to produce different starting orientations, such as a vertical Ba-O dimer, a dimer tilted at an angle, or a horizontal dimer. The most stable geometry consists of a Ba-O bent dimer with the Ba bonded to two O atoms in the bridge arrangement previously depicted in Fig. 6, with the O atom making a Sc-O-Ba bond. This bent Ba-O geometry was used as a starting point for higher Ba-O coverages on the (011) surface. An analogous procedure was also performed for Ba-O on the (111) surface. Figure 8 shows fully relaxed Ba-O structures on $Sc_2O_3$ (011) and (111) for various coverages to illustrate the relaxed adsorbate geometry. For coverages of 3 Ba-O (and higher) on the (011) surface and 6 Ba-O (and higher) on the (111) surface, the adsorbed Ba and O atoms connected to form an O-Ba-O-Ba chain structure.

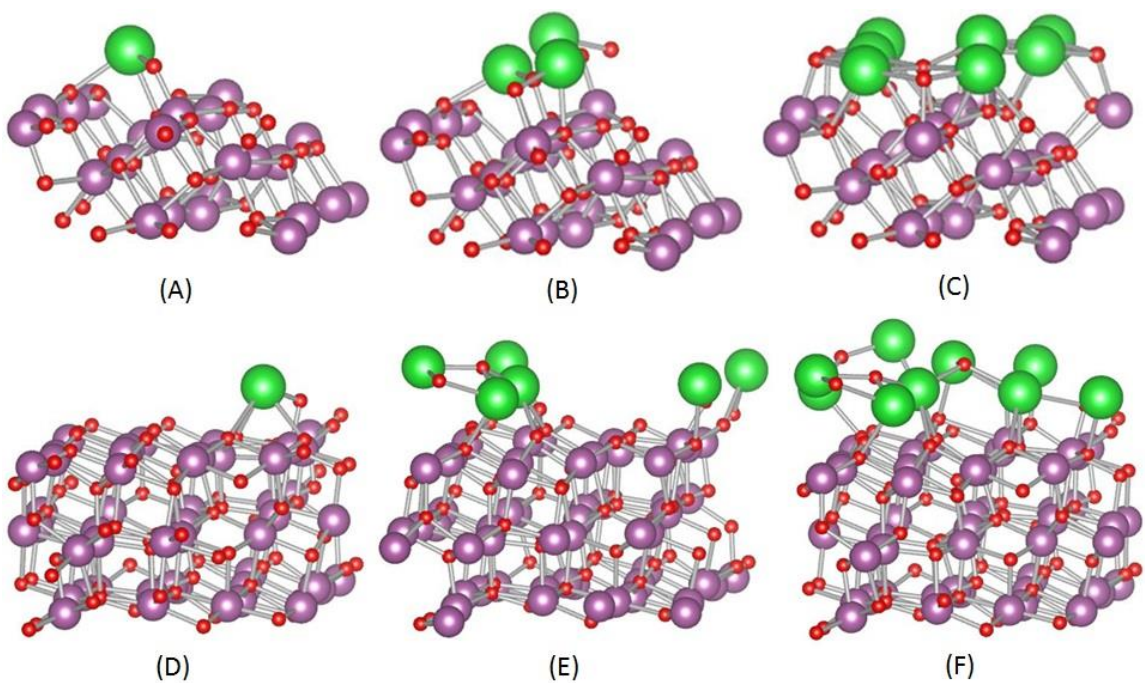

**Figure 8.** $Sc_2O_3$ (011) and (111) slabs depicting fully relaxed Ba-O dimer geometries for several example coverages. (A), (B) and (C) are coverages of 1, 3, and 7 Ba-O per (011) surface unit cell. (D), (E) and (F) are coverages of 1, 6 and 9 Ba-O per (111) surface unit cell. A stable chain geometry of O-Ba-O bonds forms at coverages between 3-6 Ba-O.

The surface barriers and surface energies for the Ba-O dimer configurations shown in Fig. 8 are given in Table 4. Figure 9 shows the surface barrier and surface energy as a function of Ba-O coverage. It can be seen for the $Sc_2O_3$ (011) surface that stable structures (relative to BaO) are realized for dimer coverages between 3 and 7 Ba-O per surface unit cell, while the 8 Ba-O coverage (1 ML BaO film) is once again unstable. In addition, all coverages of Ba-O on the (111) surface result in unstable structures. The stability of these structures versus BaO is important for two reasons. First, the relative stability of a particular Ba-O structure relative to others considered in this study serves as an indication of residence time on the $Sc_2O_3$ surface prior to evaporation. Second, bulk BaO is also considered a candidate structure on the $Sc_2O_3$ emitting surface, and any structures that are stable versus BaO (i.e. the structures below the dotted line in Fig. 8(c)) favor the formation of a partial monolayer of BaO on the $Sc_2O_3$ surface. The combination of the instability of Ba-O adsorption of all coverages on the (111) surface coupled with the surface barriers of 2.05 eV and higher indicate that Ba-O adsorption on the (111) surface is not a plausible explanation for the low surface emission barriers of scandate cathodes. Because of these facts regarding the (111) surface, we focus our discussion on Ba-O adsorption on the (011) surface. In general, our results show that the partial BaO monolayer on (011) causes the greatest surface barrier reduction, and is also stable versus bulk BaO. The instability of 1 ML BaO on (011) is most likely a result of the compressive strain placed on the BaO rocksalt lattice to have epitaxial matching with $Sc_2O_3$ (011). The removal of 1 Ba-O dimer provides additional space for strain relaxation, resulting in a more stable structure. Overall, the presence of extra O atoms stabilizes the adsorbed Ba on the (011) surface only. Thus, Ba-O dimers are more stable than Ba atoms on the (011) termination of $Sc_2O_3$, and would be expected to reside on the emitting surface for a longer period of time during operation prior to evaporation.

**Table 4.** Calculated surface barriers and surface energies for Ba-O dimer arrangements on the (011) and (111) surfaces. $\Phi_1$ and $\Phi_2$ are the surface barriers for the Ba-O containing and bare sides of the surface slab, respectively, and are for p-type conditions ($E_{Fermi}$ positioned at the VBM). The slabs which have zero Ba-O

coverage serve as the reference slabs for calculating the $\Delta\Phi_{Fermi}$ term. A coverage of 8 Ba-O dimers on the (011) surface is equivalent to a strained, single monolayer (ML) of rocksalt BaO (011).

| Termination | Coverage (dimers/unit cell) | $\Phi_1$ (eV) | $\Phi_2$ (eV) | $\Delta\Phi_{dipole}$ (eV) | $\Delta\Phi_{Fermi}$ (eV) | $\gamma$ (eV/Å$^2$) | $E_{ads}^{Ba-O}$ (eV/Ba-O) |
|---|---|---|---|---|---|---|---|
| (011) | 0 (ref) | 5.16 | 5.18 | -0.02 | n/a | 0.077 | n/a |
| | 1 | 4.02 | 4.46 | -0.44 | -0.72 | 0.080 | 0.76 |
| | 3 | 2.65 | 4.53 | -1.88 | -0.65 | 0.075 | -0.19 |
| | 6 | 2.55 | 4.73 | -2.18 | -0.45 | 0.074 | -0.15 |
| | 7 | 1.21 | 4.10 | -2.89 | -1.08 | 0.071 | -0.25 |
| | 8 (1 ML BaO) | 1.55 | 5.16 | -3.61 | -0.02 | 0.079 | 0.06 |
| (111) | 0 (ref) | 5.92 | 5.92 | 0 | n/a | 0.051 | n/a |
| | 1 | 4.48 | 5.38 | -0.90 | -0.54 | 0.052 | 0.24 |
| | 3 | 3.37 | 5.57 | -2.20 | -0.35 | 0.058 | 0.85 |
| | 6 | 2.90 | 5.95 | -3.05 | 0.03 | 0.055 | 0.27 |
| | 9 | 2.05 | 3.77 | -1.72 | -2.15 | 0.069 | 0.77 |

In contrast to atomic Ba on Sc$_2$O$_3$, the surface emission barrier lowering with Ba-O on both Sc$_2$O$_3$ (011) and (111) is due both to dipole and doping effects. Meanwhile, the chain structures produced for coverages of 3 to 7 Ba-O dimers per unit cell were calculated to be stable relative to the BaO reference state. In particular, a coverage of 7 Ba-O dimers on the (011) surface exhibited the lowest calculated surface barrier, equal to 1.21 eV and was also the most stable structure examined in this study. *This identifies the 7-BaO-dimers-on-(011) Sc$_2$O$_3$ configuration as an excellent and highly probable explanation for the observed enhanced emission of barium-impregnated-scandate cathodes.* Interestingly, the density of states for this 7 Ba-O on Sc$_2$O$_3$ (011) structure exhibited no surface bandgap, i.e. the surface metallizes. A decomposition of the density of states by row of the surface slab showed that for the top layer of adsorbed Ba and O atoms, $E_{Fermi}$ is in the conduction band, and the significant bandgap shrinkage is a result of the hybridization of states between the Sc$_2$O$_3$ slab and the adsorbed Ba-O. This surface metallization may enhance the conductivity of electrons near the surface of Sc$_2$O$_3$ that contains partial monolayers of adsorbed BaO and also aid in the enhanced emission observed in experiments.

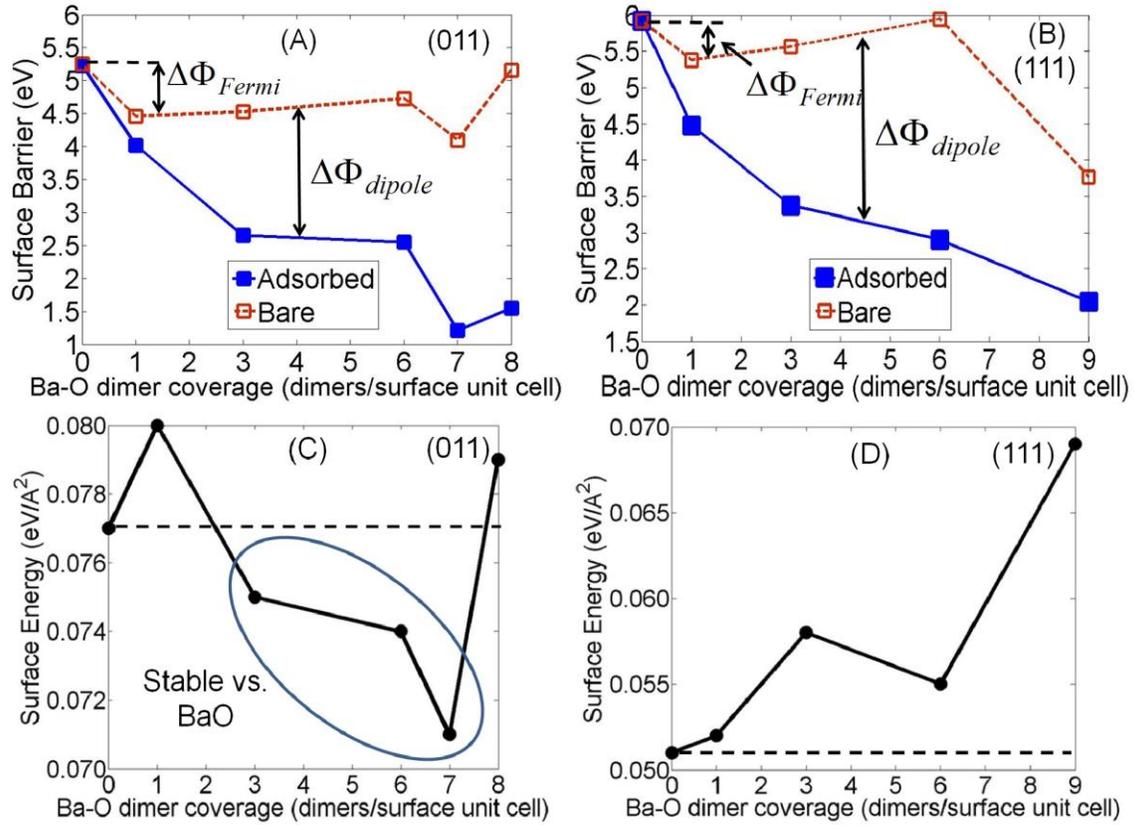

**Figure 9.** Surface barrier as a function of Ba-O coverage for the (011) termination (A) and (111) termination (B). Solid blue lines are surface barriers of the Ba-O-terminated surface while dashed red lines are the surface barriers of the opposing bare surfaces. Surface energy as a function of Ba-O coverage for the (011) termination (C) and (111) termination (D) are plotted. The dashed lines in (C) and (D) mark the surface energy of the bare reference slabs for comparison. Stable dimer coverages (relative to BaO) on (011) are circled, characterized by surface energies which are lower than the bare reference (011) surface energy.

**D. Effect of n-type bulk doping on surface barrier.** It is important to better understand the effects of doping $Sc_2O_3$ on the surface barrier, both because the $Sc_2O_3$ may be n-type in working cathodes and because doping effectively occurs for some surface species studied here (e.g., Ba surface species in Section IIIB). In this section we qualitatively investigate the effect of n-type bulk doping on the surface barrier. Here we consider the effect of bulk Li doping on the $Sc_2O_3$ (011) surfaces containing the same Ba-O coverages investigated in the previous section. Intuitively, doping the bulk of an initially pristine semiconductor n-type should raise $E_{Fermi}$, thus lowering the surface barrier.

To investigate how n-type bulk doping affects the surface barrier, we use an interstitial Li atom as a bulk dopant. Li was used not because it represents a real impurity in $Sc_2O_3$ materials, but because it readily donates its electron and serves as an approximation for heavily n-type conditions. Table 5 contains the calculated surface barriers for Li-doped $Sc_2O_3$ (011) surfaces with various Ba-O dimer coverages, along with the undoped bare surface and undoped surface with 7 Ba-O dimers as references for comparison, while all undoped Ba-O surface barriers are in Table 4). Overall, we see an expected behavior of a lower surface barrier for the bare (011) surface, lowered from 5.16 to 1.77 eV on one surface face and from 5.18 to 1.69 eV on the opposing surface face. The decrease of the surface barrier in the case of the bare (011) surface is 3.4-3.5 eV, which can be explained by the shifting of $E_{Fermi}$ of the surface slab by approximately the surface bandgap, which with HSE for the (011) surface is 3.5 eV. These impacts of the doping on the DOS can be seen in Fig. 10. The bandgap in the vicinity of the surface differs from that of the bulk (bulk HSE gap is 5.8 eV, (011) surface HSE gap is 3.5 eV), and as a result one should not generally expect that raising $E_{Fermi}$ by a certain value in the bulk (for example, by purposeful n-type doping) will lead to an identically sized decrease in the surface barrier.

**Table 5.** Calculated surface barriers for several Ba-O dimer coverages on the (011) Li doped surface. The data for 0 and 7 Ba-O dimers on the undoped (011) surface are repeated for ease of comparison. For nonzero Ba-O coverages, $\Phi_1$ and $\Phi_2$ are the surface barriers for the Ba-O containing and bare sides of the surface slab, respectively. The surface barriers for undoped cases are for pristine conditions and those for the Li doped cases are under heavily n-type conditions.

|   | Coverage (dimers/unit cell) | $\Phi_1$ (eV) | $\Phi_2$ (eV) | $\Delta\Phi_{dipole}$ (eV) | $\Delta\Phi_{Fermi}$ (eV) |
|---|---|---|---|---|---|
| (011) undoped | 0 | 5.16 | 5.18 | -0.02 | n/a |
|  | 7 | 1.21 | 4.10 | -2.89 | -1.08 |
| (011) Li doped | 0 | 1.77 | 1.69 | 0.08 | -3.49 |
|  | 1 | 1.68 | 1.68 | 0.00 | -3.50 |
|  | 3 | 1.68 | 1.68 | 0.00 | -3.50 |
|  | 6 | 1.77 | 1.65 | 0.12 | -3.53 |
|  | 7 | 1.83 | 2.27 | -0.44 | -2.91 |
|  | 8 | 1.99 | 2.13 | -0.14 | -3.05 |

In Section IIIC, when the pristine $Sc_2O_3$ system was studied and Ba-O coverage was increased, the surface barrier was lowered by a combination of surface dipoles and electron doping. Interestingly, under the heavy n-type conditions Li doping provides, the surface barrier is nearly unchanged between 0 and 6 Ba-O dimers per surface unit cell. From Table 5, the calculated surface barrier is between 1.68-1.77 eV for these coverages, and the surface barrier lowering is due nearly entirely from electron doping (similar to atomic Ba adsorption from Section IIIB). For the highest coverages of 7 and 8 Ba-O dimers, there is also a small dipole contribution; however this dipole contribution has been significantly lessened by the presence of the Li dopant (compare, for example, the surface barriers and dipole components of the doped and undoped surfaces containing 7 Ba-O dimers in Table 5). Interestingly, when Li doping is included with high Ba-O coverages of 7 and 8 dimers per surface unit cell, the surface barrier actually increases compared to the undoped surface of Section IIIC, which is a counterintuitive result. For the remainder of this discussion, we focus on the 7 Ba-O coverage as this surface arrangement yielded the lowest calculated surface barrier from Section IIIC. Figure 10 is a plot of the density of states of the top atomic row of the (011) surface containing 7 Ba-O, and shows an upward shift of $E_{Fermi}$ with respect to the undoped case, which would be expected to lower the surface barrier. However, from performing Bader charge analysis it was found that for the undoped $Sc_2O_3$ (011) 7 Ba-O terminated slab, the adsorbed Ba-O dimers transferred roughly 0.52 more electrons per surface unit cell to the $Sc_2O_3$ slab layers than when the $Sc_2O_3$ was doped with Li. By inspecting the values of $\Delta\Phi_{dipole}$ and $\Delta\Phi_{Fermi}$ in Table 5, one can see that there is a trade-off of surface barrier lowering due to the surface dipole effect and due to doping effects. When both the Li doping and adsorbed Ba-O are present together, the dipole effect is significantly reduced due to less charge transfer from the surface Ba-O species, and the doping component is raised slightly and is now due to both the Li impurity and the surface Ba-O. The overall result is that when there is nearly a full monolayer of Ba-O coverage, the surface barrier is higher when the system is heavily n-type and lower when it is undoped. In addition, the results of adding Li demonstrate that the dipole and doping components are coupled, i.e., doping the system can change the surface dipole. One cannot simply maximize both the surface dipole and a separate doping contribution in order to maximally lower the surface barrier.

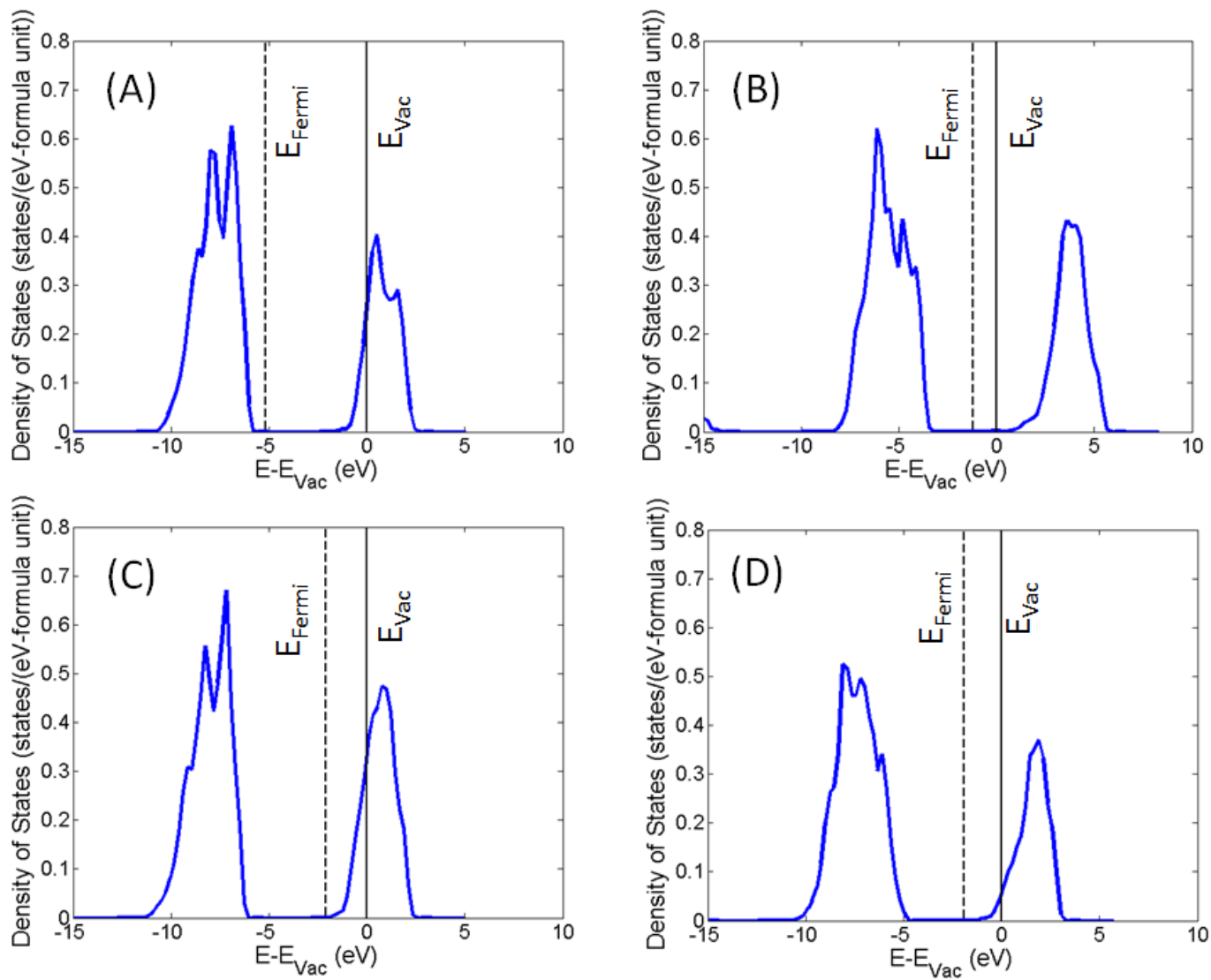

**Figure 10.** Total densities of states for the top row of Sc-O atoms for undoped (A) bare $Sc_2O_3$ (011) slab, (B) $Sc_2O_3$ slab with 7 Ba-O on surface, (C) bare $Sc_2O_3$ (011) slab with Li-doping, and (D) $Sc_2O_3$ slab with 7 Ba-O dimers, with Li doping. The black solid and dashed lines on the plots mark the vacuum and Fermi levels, respectively. Both cases show an upward shift of Fermi level, though the doping shift is much greater in the case of the bare surface (A, C).

This brief investigation on the effects of n-type doping has a few implications with regard to experimental cathodes. First of all, it is probable that the real scandate cathode system is n-type so that electron conduction is high enough to supply the necessary emission current. We have found that under very heavy n-type conditions ( $E_{Fermi}$ in the conduction band) that when the $Sc_2O_3$ (011) surface is bare, there is a large surface barrier reduction, while if the $Sc_2O_3$ surface contains high Ba-O dimer coverages near one monolayer, the surface barrier may actually be raised since the surface dipole is reduced. At first glance our argument appears contradictory since we have claimed that the system should be n-type for ample conduction but also contain surface adsorbed Ba-O for surface barrier lowering. However, it is not probable that the real scandate system is so heavily n-type that $E_{Fermi}$ is in the conduction band, and thus our results with Li doping represent an extreme case where if the system is doped this heavily one may see surface barrier raising when Ba-O is present on the surface. We believe it is most probable that the system is weakly n-type and $E_{Fermi}$ lies within the bandgap, and the doping and surface adsorbed Ba-O both act to lower the surface barrier.

Overall, the analysis of different Ba adsorption mechanisms on $Sc_2O_3$ surfaces demonstrates how one may tune a material's surface barrier. Charge transfer is critically important, and in general charge must flow from the adsorbate to the substrate if the surface barrier is to be lowered. A dipole or doping contribution may dominate depending on the specific details of the surface species present and the Fermi level of the initial material (e.g. as set by additional bulk dopants). Doping the bulk of the material has a direct effect on the surface barrier through changing $E_{Fermi}$, but also an indirect effect through changing $E_{Fermi}$ altering the magnitude of the surface dipole. However, it is not clear how to tell *a priori* how the charge doping from impurities affects the magnitude of the surface dipole or the magnitude of the doping contribution due to adsorbed surface species. The net outcome involves a potentially complicated interplay between the surface species and bulk dopants to set $E_{Fermi}$, which in turn affects the surface dipole since a different quantity of charge may be transferred to the material from surface adsorbates once bulk dopants are also present. In this way, there is a significant coupling that inhibits surface barrier tuning in semiconductors when both bulk dopants and surface dipole species are present. If the material is doped n-type and $E_{Fermi}$ is pushed higher, this makes it energetically more difficult for the surface species to form surface barrier-lowering dipoles, as these dipoles require charge transfer of electrons to the vicinity of the surface. Therefore, there is a tradeoff between bulk doping and surface dipole formation when realizing the maximum possible change in surface barrier of a semiconductor.

## IV. CONCLUSIONS

The surface stability and surface barrier characteristics of the low index surfaces of $Sc_2O_3$ were investigated using DFT-based methods with the GGA and hybrid HSE functionals in an effort to explain the superior emission characteristics observed from scandate cathodes in experiment. Bare (001), (011), and (111) $Sc_2O_3$ surfaces and (011) and (111) surfaces with adsorbed atomic Ba and Ba-O dimers were explored. For the bare $Sc_2O_3$ surfaces, it was found that the order of stability, calculated by the surface energy, is (from most to least stable): $\gamma(111) < \gamma(011) < \gamma(001)$. The (001) surface terminations are polar and very unstable, thus the (001) terminations are not expected to function as dominant emitting surfaces. Therefore, it is expected that the (011) and (111) surfaces are the dominant emitting surfaces.

The surface barriers for bare $Sc_2O_3$ surfaces range from 5.16 eV for the (011) to 5.92 eV for the (111) and between 2.10-2.30 eV when both surfaces contain small amounts of adsorbed atomic Ba. However, atomic Ba adsorption does not produce stable structures versus BaO. In addition, adsorption of Ba-O dimers was considered, as these are likely to form due to reaction with O in the cathode operating environment, i.e. high temperatures of approximately 1200 K and low pressures of approximately $10^{-10}$ Torr. For the (111) surface, no coverage of Ba-O dimers was found to be stable versus BaO, but a number of stable configurations versus BaO were found for the (011) surface. In particular, 7 Ba-O dimers per (011) surface unit cell, was calculated to be the most stable surface adsorbate structure, both relative to all other structures considered in this study and relative to bulk BaO, indicating that adsorbed Ba-O on the $Sc_2O_3$ surface will prefer to adopt a sub-monolayer structure that will reside on the emitting surface for the longest time relative to other structures considered here prior to evaporating. This 7 Ba-O dimers per surface unit cell on the (011) surface also yielded the lowest calculated surface emission barrier of 1.21 eV. Finally, no bandgap was detectable at the surface from the density of states data for this structure due to the large number of Ba and O states from the adsorbed Ba-O dimers. From both the stability data and calculated surface barrier values, we conclude that of all structures and coverages studied, the 7-BaO-dimer-on-Sc$_2$O$_3$ is the most probable configuration, and its low 1.21 eV surface emission barrier is the closest example of what can be regarded as a true work function of Sc$_2$O$_3$ plus adsorbed Ba-O.

This 1.21 eV work function suggests that crystalline Sc$_2$O$_3$ with BaO can exhibit low enough surface barriers to be consistent with results observed in experimental thermionic cathodes. As discussed in the introduction, experimental effective work functions for scandate cathodes obtained from Miram and PWFD curves are in the range of 1.3-1.5 eV. Since these effective work functions are a complex average of the entire emitting area, it is sensible that the lowest emission surface should have a lower local work function than the total effective work function, which is what we find here. Indeed, some surface patches may be characterized by the Sc$_2$O$_3$+ Ba-O dimer coverages we calculated here, while other patches may consist of Sc$_2$O$_3$ devoid of Ba, and other patches that are just bare W. In addition, grain boundaries, voids, and defected or off-stoichiometric Sc$_2$O$_3$ crystals provide additional surface structures not considered here which also may contribute to the total measured emission current and factor into the effective work function measurement. Overall, the Sc$_2$O$_3$ structures considered in this study coupled with the physics of Ba interaction on these surfaces provides a consistent picture of work function reduction in scandate emitters.

The effect of bulk n-type doping on the surface barrier was briefly investigated on the bare Sc$_2$O$_3$ (011) surface and the (011) surface containing several Ba-O dimer coverages. In all cases, the interstitial Li used as an electron donor doped electrons into the system, and raised the position of $E_{Fermi}$. When no surface species were present, this led to a straightforward lowering of the surface barrier by doping. Even when Ba-O dimers were adsorbed up to a coverage of 6 Ba-O per (011) surface unit cell, the surface barrier lowering was due entirely to electron doping and essentially no dipole formation was observed. When a significant surface dipole was present with 7 and 8 adsorbed Ba-O, the surface barrier actually increased. This effect occurred because the electron doping by Li made it so that the adsorbed Ba-O could not transfer as much charge to the $Sc_2O_3$ compared to when it is undoped, thus diminishing the surface dipole component. Overall, the surface dipole and doping contributions to surface barrier reduction are coupled and their effects cannot be added independently. This fact imposes a potential limitation of work function tuning in semiconducting materials. If one introduces bulk dopants into a material that makes the material more n-type, then modification of the surface with dipole species has a diminishing effect, as $E_{Fermi}$ of the entire material is higher, so even charge transfer by dipoles that is restricted to just the surface region becomes energetically more difficult and is suppressed since $E_{Fermi}$ of the entire system has been raised by the bulk dopant species.

The topics discussed in this work have an impact beyond those interested in using semiconducting materials for electron emission into vacuum. Doping and surface dipole effects are two key ingredients which determine how surface barriers may be modified. Surface barrier engineering (more commonly referred to as "work function engineering") is critical in a wide range of materials applications where electronic transport across interfaces to either vacuum or another material is of interest. Examples include transistors, solar cells, solid state memory, electron emitters, and even catalysts.

## AUTHOR INFORMATION

**Corresponding Authors**

*Tel.: (608) 265-5879. E-mail: ddmorgan@wisc.edu

*Tel.: (608) 890-0804. E-mail: jhbooske@engr.wisc.edu

## ACKNOWLEDGEMENTS

This work was supported by the US Air Force Office of Scientific Research through grants No. FA9550-08-0052 and No. FA9550-11-0299. Computational support was provided by a NERSC allocation from the Center for Nanophase Materials Sciences (CNMS) at Oak Ridge National Laboratory under grant number CNMS2013-292.

TOC Figure (2 in. x 2 in.):

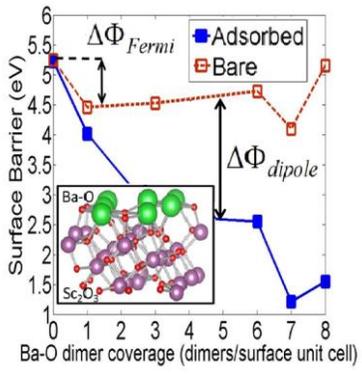